\documentclass[preprint,5p,twocolumn,showpacs,epsfig,10pt]{elsarticle}
\usepackage{color}

\RequirePackage{amsfonts}
\usepackage{colordvi}

\usepackage{graphicx,amssymb,epsfig,epsf,amsmath}

\newcommand{\rd}{{\rm d}}

\newcommand{\be}{\begin{equation}}
\newcommand{\ee}{\end{equation}}
\newcommand{\bea}{\begin{eqnarray}}
\newcommand{\eea}{\end{eqnarray}}

\journal{Physics Letters A}

\begin{document}

\begin{frontmatter}

\title{Structural and quantum properties of van der Waals cluster near the unitary regime}

\author{M. L. Lekala$^1$, B. Chakrabarti$^{2}$, S. K. Haldar$^{3,4}$, R.Roy$^{2}$, G. J. Rampho$^{1}$}

%
\address{
$^{1}$Department of Physics, University of South Africa,
P.O.Box 392, Pretoria 0003, South Africa.\\
$^{2}$ Department of Physics, 
Presidency University, 86/1 College Street, Kolkata 700 073, India. \\
$^{3}$Theoretical Physics Division, Physical Research Laboratory,
Navrangpura, Ahmedabad 380009, India.\\
$^4$Department of Physics, University of Haifa, Haifa, Israel.
}

\begin{abstract}
We study the structural and several quantum properties of three-dimensional bosonic cluster interacting through van der Waals potential at 
large scattering length. We use Faddeev-type decomposition of 
the many-body wave function which includes all possible two-body correlations. At large scattering length, we observe spatially extended states which exhibit the exponential dependence on the state number. The cluster ground state energy shows universal nature at large negative scattering length.
We also find the existence of generalized Tjon lines for $N$- body clusters. Signature of universal behaviour of weakly 
bound clusters can be observed in experiments of ultracold Bose gases. We also study the spectral statistics of the system. We calculate both the short-range fluctuation and long-range correlation and observe semi-Poisson distribution which interpolates the Gaussian Orthogonal Ensemble (GOE) and Poisson statistics of random matrix theory. It indicates that the van der Waal cluster near the unitary becomes highly complex and correlated. However additional study of $P(r)$ distribution (without unfolding of energy spectrum) reveals the possibility of chaos for larger cluster.
\end{abstract}
\end{frontmatter}
\section{Introduction}

Computation of energy levels and the study of various structural and quantum properties of several rare gas clusters is a long standing problem in many-body physics and several remarkable works have been published in this direction~\cite{Leit,Rich,Mush,Blum,Blum1,Blum2,Blum3}. The properties of these clusters are mainly calculated using variational Monte Carlo methods. It has been observed that in contrast to helium, the rare gases have more classical behaviour. The interaction potential is generally taken as the Lennard-Jones potential. The energetics and structural properties of super borromean $N$-body clusters has also been reported recently~\cite{Blum4}.

Our present study mainly involves the characterization of universal properties of bosonic clusters in the unitary regime and interacting with van der Waals potential. The effective interatomic interaction at ultracold temperature can be essentially tuned to any desired value by utilizing Feshbach resonances. By changing the external magnetic field the evidence for the formation of Efimov trimer states have been reported~\cite{kra,huc}. 
The study of universalities in few-body quantum system $(N>3)$ is not straightforward. There are several studies in this direction which predicts the universality of these systems~\cite{bra,sor,ste,had,smi,vig,jia,don,esr,nie}, though predictions and conclusions made in these works are qualitatively similar, quantitative difference exist~\cite{ste1,ham,pol,blu,gip,yam}. This is to be noted that energy level statistics and spectral analysis of van der Waals clusters are reported earlier by our group~\cite{Pankaj,PRA2014} . However the earlier calculations consider diffuse cluster and weakly interacting trapped bosons. Whereas the present study considers the atomic cluster at very large scattering length when the system becomes highly correlated and complex. The comparison with the diffuse cluster is made in a separate section later. 

By using the Feshbach resonance the two-body scattering 
length $a_{s}$ is tuned to very large values. The unitary regime is characterized 
by simple universal laws. For weakly interacting dilute Bose gas, the gas like state becomes unstable as $a_{s}$ increases~\cite{don}. However in quantum few-body systems it leads to different concept of universality. 
Universalities appear when the attractive two-body interaction is such that the scattering 
length is much larger than the range of the interaction. 
Under such condition, a series of weakly 
bound and spatially extended states appear in the system.
Although the behaviour of ultracold Fermi 
gas is well understood, the exhaustive study of bosonic system with large scattering length 
are few. Helium trimer $^{4}He_{3}$ is a well studied quantum three-body system in this direction~\cite{esr,nie}.

In this work we consider few-bosonic clusters of $^{85}Rb$ atoms interacting with van der Waals interaction. In some recent experiments it has been revealed that the range of the interaction between atoms is typically the van der Waals length characterized by $r_{vdw}= \frac{1}{2} \left( \frac{mC_{6}}{\hbar^{2}} \right)^{\frac{1}{4}}$, which is associated with the $-\frac{C_{6}}{r^{6}}$ tail~\cite{chin,naidon}. To characterize this delicate system we prescribe two-body correlated basis function for the many-body cluster. At large scattering length the ground state energy exhibits the universality for few atoms ($N$ upto 7) in the cluster. A series of spatially extended states are observed which exhibit exponential dependence on the state number. We also find the existence of generalized Tjon lines for $N$-body clusters.

The study of energy level statistics of such complex clusters is another important area of our present study. We able to calculate the full energy spectrum for $N=7$ cluster and calculate nearest-neighbour level spacing distribution $P(s)$ and $\bigtriangleup_{3}$ statistics. Our numerical results strongly resembles the semi-Poisson distribution for the lower levels which interpolates the GOE (Gaussian Orthogonal Ensemble) and Poisson statistics. We also present the $P(r)$ distribution of the ratio of the consecutive level spacing $(r)$ and the average $<\tilde{r}>$ which is presently considered as the most useful statistical measures to distinguish order and chaos in the energy levels.

The paper is organized as follows. In Sec.~2 we discuss the many-body Hamiltonian and numerical calculation. 
Sec.~3 considers the results and exhibit the signature of universal cluster states. Comparison is made with diffuse cluster. It also present calculation of energy level statistics. Sec.~IV concludes with a summary.

\section{Many-body Hamiltonian and numerical calculations}

We approximately solve the many-body Schr\"odinger equation by Potential harmonic expansion method (PHEM). 
We have successfully applied PHEM to study different properties of Bose Einstein condensate~\cite{Tapan, Das, Kundu} 
and atomic clusters~\cite{TKD, Pankaj, PRA2014}. 
The method has been described in detail in our earlier works~\cite{Tapan, Das, Kundu}. We briefly describe the method below for interested readers.

We consider a system of $N=(A+1)$ $^{85}$Rb atoms, each of mass $m$ and interacting via two-body potential. 
The Hamiltonian of the system is given by
\begin{equation}
H=-\frac{\hbar^2}{2m}\sum_{i=1}^{N} \nabla_{i}^{2} 
+\displaystyle{\sum_{i,j>i}^{N}} V(\vec{r}_{i}-\vec{r}_{j})\cdot
\label{eq.H}
\end{equation}
Here $V(\vec{r}_{i}-\vec{r}_{j})$ is the two-body potential and 
$\vec{r}_{i}$ is the position vector of the $i$th particle. It is usual practice to decompose the motion of a many-body system into  
the motion of the center of mass where the center of mass coordinate is $\vec{R}=\frac{1}{N}\sum_{i=1}^{N} \vec{r}_{i}$ 
and the relative motion of the particles in center of mass frame. For atomic clusters,
the center of mass behaves like a free particle in laboratory frame and we set its energy zero. 
Hence, we can eliminate the center of mass motion by using standard 
Jacobi coordinates, defined as~\cite{Ballot, Fabre, MFabre}
\begin{equation}
\vec{\zeta}_{i}=\sqrt{\frac{2i}{i+1}}(\vec{r}_{i+1}-
\frac{1}{i}\sum_{j=1}^{i} \vec{r}_j) \hspace*{.5cm}
 (i=1,\cdots,A),
 \label{eq.jacobi}
\end{equation}
and obtain the Hamiltonian for the relative motion of the atoms
\begin{equation}
H=-\frac{\hbar^{2}}{m}\sum_{i=1}^{A} 
\nabla_{\zeta_{i}}^{2} + V_{int}
(\vec{\zeta}_{1}, ..., \vec{\zeta}_{A})\hspace*{.1cm}\cdot
\label{eq.relative-H}
\end{equation} 
Here $V_{int}(\vec{\zeta}_{1}, ..., \vec{\zeta}_{A})$ is the sum of all pair-wise interactions expressed in terms of Jacobi coordinates. 
The Hyperspherical harmonic expansion method (HHEM) 
is an {\it ab-initio} complete many-body approach and includes all possible correlations. 
The hyperspherical variables are constituted by the hyperradius $r = \sqrt{\sum_{i=1}^A\zeta_{i}^{2}}$ and $(3A-1)$ 
hyperangular variables which are comprised of $2A$ spherical polar angles $(\vartheta_j,\varphi_j; \ j=1,\cdots,A)$ 
associated with $A$ Jacobi vectors and $(A-1)$ hyperangles $(\phi_2,\phi_3,\cdots,\phi_A)$ given by their lengths. 
However the calculation of potential matrix elements of all pairwise potentials becomes a formidable task and the convergence rate of 
the hyperspherical harmonic expansion becomes extremely slow for $N > 3$, due to rapidly increasing degeneracy of the basis. 
Thus HHEM is not suitable for the description of large and complex atomic clusters. 
However we may assume that only two-body correlation and pairwise interaction are present and the total wave function $\Psi$ can be decomposed into two-body Faddeev component 
for the interacting $(ij)$ pair as 
\begin{equation}
\Psi=\sum_{i,j>i}^{N}\phi_{ij}(\vec{r}_{ij},r)\hspace*{.1cm}\cdot
\label{eq.Faddeev-comp}
\end{equation}
$\phi_{ij}$ is a function of two-body 
separation ($\vec{r}_{ij}$) and the global 
$r$ only. Therefore for each of the $N(N-1)/2$ interacting pair of a $N$ particle system, the active degrees of freedom is effectively reduced to 
only four, {\it viz.}, $\vec{r}_{ij}$ and $r$ and the remaining irrelevant degrees of freedom are frozen. 
Since $\Psi$ is decomposed into all possible interacting pair Faddeev components, 
{\it all two-body correlations} are included. Thus the physical picture for a given Faddeev component is that when two particles interact, 
the rest of the particles behave as inert spectators. The two-body correlation enters through the two-body expansion basis and as $\phi_{ij}$ is symmetric under the exchange operator $P_{ij}$, the Faddeev equation can be written as 
\begin{equation}
\left[T-E\right]\phi_{ij}
=-V(\vec{r}_{ij})\sum_{kl>k}^{N}\phi_{kl}
\label{eq.Faddeev-eqn}
\end{equation}
where $T = -\frac{\hbar^2}{m} \displaystyle{\sum_{i=1}^A} \nabla_{\zeta_{i}}^{2}$ is the total kinetic energy operator. 
Applying the operator  $\sum_{i,j>i}$ on both sides of Eq.~(\ref{eq.Faddeev-eqn}), we get back the original Schr\"odinger equation.
Since we assume that when ($ij$) pair interacts the rest 
of the bosons are inert spectators, the total hyperangular momentum 
and the orbital angular momentum of the whole system 
is contributed by the interacting pair only. 
The  $(ij)$th Faddeev 
component is then expanded in the subset of hyperspherical harmonics, which we call as potential harmonic (PH) basis as 
\begin{equation}
\phi_{ij}(\vec{r}_{ij},r)
=r^{-(\frac{3A-1}{2})}\sum_{K}{\mathcal P}_{2K+l}^{lm}
(\Omega_{A}^{ij})u_{K}^{l}(r) \hspace*{.1cm}\cdot
\label{eq.expansion}
\end{equation}
${\mathcal P}_{2K+l}^{lm}(\Omega_{A}^{ij})$ is called the PH. It 
has an analytic expression:
\begin{equation}
{\mathcal P}_{2K+l}^{l,m} (\Omega_{A}^{(ij)}) =
Y_{lm}(\omega_{ij})\hspace*{.1cm} 
^{(A)}P_{2K+l}^{l,0}(\phi) {\mathcal Y}_{0}(D-3) ;\hspace*{.5cm}D=3A ,
\label{eq.PH}
\end{equation}
${\mathcal Y}_{0}(D-3)$ is the HH of order zero in 
the $(3A-3)$ dimensional space.
The global hyperradius $r$ is further defined as $r^{2}$ = $r_{ij}^{2}+\rho_{ij}^{2}$, where $r_{ij}$ is the separation between $(ij)$ interacting pair and $\rho_{ij}$ is basically the global size of the remaining noninteracting bosons. As the angular momentum contribution from the noninteracting $(A-1)$ bosons is zero, the $3A$ quantum number of HH is now reduced to only four as energy $E$, orbital angular momentum quantum number $l$, azimuthal qunatum number $m$, grand orbital quantum number $2K+l$ for any $N$. We substitute Eq.(4) in eq.(5) and takes a projection on a particularPH basis and obtain a set of coupled differential equation in the partial wave $U_{Kl}(r)$.
\begin{equation}
\begin{array}{cl}
\Big[\displaystyle{-\frac{\hbar^{2}}{m} \frac{\rd^{2}}{\rd r^{2}} + \frac{\hbar^{2}}{mr^{2}}
\{ {\cal L}({\cal L}+1) 
+ 4K(K+\alpha+\beta+1)\}} &\\
-E\Big]U_{Kl}(r) +\displaystyle{\sum_{K^{\prime}}}f_{Kl}V_{KK^{\prime}}(r)
f_{K^{\prime}l}
U_{K^{\prime}l}(r) = 0&\\
\end{array}
\label{eq.cde}
\end{equation}\\
where ${\mathcal L}=l+\frac{3N-6}{2}$, $U_{Kl}=f_{Kl}u_{K}^{l}(r)$, 
$\alpha=\frac{3N-8}{2}$ and $\beta=l+1/2$.\\
$f_{Kl}$ is a constant and represents the overlap of the PH for
interacting partition with the sum of PHs corresponding  to all 
partitions~\cite{MFabre}.
The potential matrix element $V_{KK^{\prime}}(r)$ is given by

\begin{eqnarray}
\begin{split}
V_{KK^{\prime}}(r)=\int P_{2K+l}^{lm^*}(\Omega_{A}^{ij})V\left(r_{ij}\right)
P_{2K^{\prime}+1}^{lm}(\Omega_{A}^{ij}) \rd\Omega_{A}^{ij} \\
=\left( h_K^{\alpha \beta} h_{K^\prime}^{\alpha \beta}\right)\int_{-1}^{1}P_K^{\alpha \beta}(z) V\left(r\sqrt{\frac{1+z}{2}}\right) \\
\times P_{K^\prime}^{\alpha \beta}(z) \omega_l(z) dz
\hspace*{.1cm}\cdot
\end{split}
\label{eq.potmat}
\end{eqnarray}

We do not require the additional short-range correlation function $\eta(r_{ij})$ as mentioned in Ref.~\cite{PRA2014}.
\section{Results}
\label{3}
\subsection{Universal Cluster states}

It is already pointed out that the universal properties of ultracold dilute atomic gas in the unitary regime 
is characterized when the two-body scattering length $a_{s}$ is tuned to very large values by using the Feshbach resonance. The unitary regime is characterized by simple universal laws. 

\begin{figure}[hbpt]
\vspace{-10pt}
\centerline{
\hspace{-3.3mm}
\rotatebox{0}{\epsfxsize=8cm\epsfbox{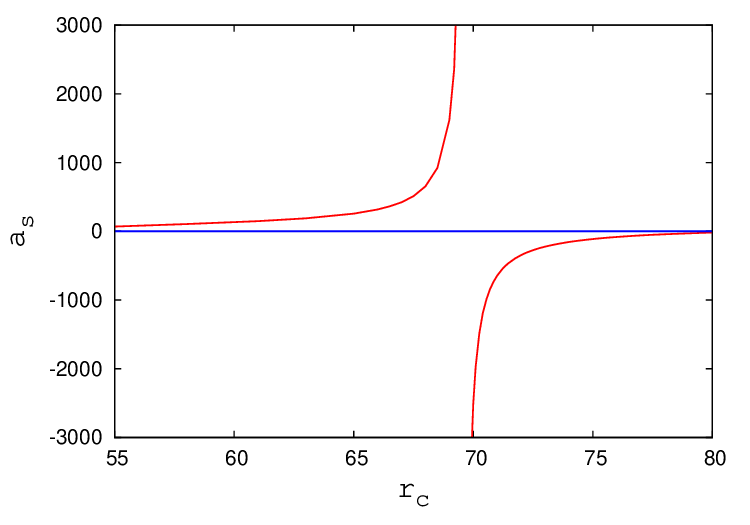}}}
\caption{Plot of zero energy scattering length $a_s$ (in Bohr) as a function of $r_c$ (in Bohr). Here only one branch corresponding to zero node
in two-body wave function is shown. Blue horizontal line shows the $a_s=0$.}
\label{fig.as}
\end{figure}

Although the unitary Fermi gas has been largely investigated both experimentally and theoretically~\cite{gip}, the bosonic unitary 
regime is a formidable challenge in the many-body theories. Even though the range of the interaction is small compared with 
the particle separation, interatomic correlations are very important and the standard mean-field theories are inadequate.\\
The interaction strength of sufficiently dilute atomic cloud is parameterized by a single parameter-the $s$-wave scattering length. 
However for our present study to explore the generic behaviour near the unitary, we consider the 
van der Waals potential characterized by two parameters: the cutoff radius of the repulsive hard core $r_{c}$ and 
the strength of the long-range tail $C_{6}$. Thus keeping $C_{6}$ fixed, it is possible to tune the value of $r_{c}$. 
Solving the two-body Schr\"odinger equation it is possible to calculate the scattering length for each choice of $r_{c}$. 
We solve the zero-energy two-body Schr\"odinger equation for the two-body wave function $\eta(r_{ij})$ as 
\begin{equation}
-\frac{\hbar^2}{m}\frac{1}{r_{ij}^2}\frac{\rd}{\rd r_{ij}}\left(r_{ij}^2
\frac{\rd\eta(r_{ij})}{\rd r_{ij}}\right)+V(r_{ij})\eta(r_{ij})=0
\hspace*{.1cm}\cdot
\label{eq.tbe}
\end{equation}
Where $V(r_{ij}) = \infty$ for $r_{ij} \leq r_{c}$ and $-\frac {C_{6}}{r_{ij}^{6}}$ for $r_{ij} > r_{c}$. 
The asymptotic form of $\eta(r_{ij})$ is $C\big( 1-\frac{a_{s}}{r_{ij}} \big)$, $C$ is the normalization constant. 
The solution of two-body equation shows that the value of $a_{s}$ changes from negative to positive passing through an 
infinite discontinuity. In Fig.~\ref{fig.as}, we plot the zero-energy scattering length $a_{s}$ as a function of $r_{c}$. 
At each discontinuity, one extra node in the two-body wave function appears which corresponds to one extra two-body bound state. 
However for our present study we fix $r_{c}$ such that it corresponds to zero node in the two-body wave function. 
We impose the constraint just to avoid the formation of the molecules, otherwise when $a_{s}$ sufficiently increases, 
the rate of three-body collisions will increase which deplete the density by forming molecules. For the present calculation we keep $C_6 = 1.0295 \times 10^{9} cm^{-1}a_0^6$ (which corresponds to $Rb$ atom) and tune $r_c$ gradually to achieve the unitary regime. For the present choice of $C_6$ parameter, the unitary regime is achieved at $r_c = 69.67 a_0$. This apparently appears the cutoff at larger distance. However tuning $C_6$, one can make $r_c$ smaller and can consider the real experimental situation. \cite{author}  \\
With the above set of parameters we solve the set of coupled differential equations (CDEs) by hyperspherical adiabatic approximation (HAA)~\cite{Coelho}. 
In HAA, we assume the hyperradial motion is slow compared to the hyperangular motion. Thus the solution of the 
hyperangular motion is obtained by diagonalizing the potential matrix including the diagonal hypercentrifugal repulsion for a fixed value of $r$. 
The CDE is then decoupled approximately into a single uncoupled differential equation 
\begin{equation}
\left[-\frac{\hbar^{2}}{m}\frac{d^{2}}{dr^{2}}+\omega_{0}(r)-E_{R}
\right]\zeta_{0}(r)=0\hspace*{.1cm},
\label{eq.EAA}
\end{equation}
which is known as extreme adiabatic approximation (EAA) and the lowest eigenvalue $\omega_{0}(r)$ is the effective potential 
in which the hyperradial motion takes place. The above equation is solved to obtain the energy and wave function with appropriate 
boundary conditions on $\zeta_{0} (r)$.\\
\begin{figure}[hbpt]
\vspace{-10pt}
\centerline{
\hspace{-3.3mm}
\rotatebox{0}{\epsfxsize=8cm\epsfbox{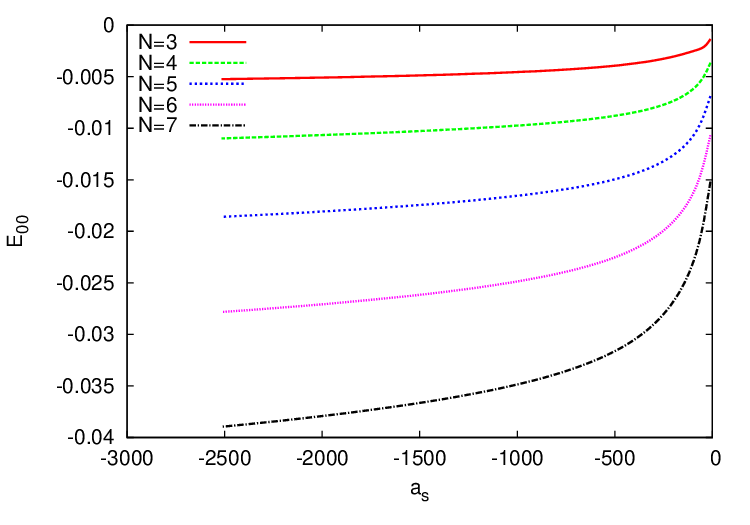}}}
\caption{Plot of ground state energy $E_{00}$ (in cm$^{-1}$) of van der Waals clusters of different cluster sizes $N$ 
as a function of $s-$ wave scattering length $a_s$ (in Bohr).}
\label{fig.gndstate}
\end{figure}
In Fig.~\ref{fig.gndstate}, we plot the calculated bosonic cluster ground state energies in the negative scattering length near the unitary for 
different cluster sizes with $N$ = 3,4,5,6,7 as a function of the scattering length $a_{s}$ which represents the universal properties 
of the bosonic cluster energy at large $|a_{s}|$. It is to be noted that the effective interaction of the bosonic cluster 
is determined by $\int V(r_{ij}) \rd^{3}r_{ij}$. With increase in particle number, the number of interacting pair $\left( \frac{N(N-1)}{2} \right)$ 
also increases and the energy becomes more negative as expected. 
We have calculated the spectrum of 
bosonic clusters $E_{n0}$ and the radii $r_{av}$ and plot them in Fig.3 and in Fig.4 respectively as a function of the state number $n$ of the negative energy states. 
In Fig.3, we observe that for each of the $N$-body systems there is a series of bound states with exceedingly small energies. 
It is seen that these series of states show exponential dependence upon the state number as $E_{n0}$ $\propto$ $e^{-B_{N}n}$. 
The exponential fits give the numbers as $B_{4}$ =0.448, $B_{5}$ = 0.278, $B_{6}$ = 0.198, $B_{7}$ = 0.149. Whereas in Fig.~\ref{fig.rav}, 
we observe that the spatial extension of the states is much larger than the interaction range and the r. m. s. radii  
are well reproduced with the exponential $R_{n}$ $\propto$ $e^{C_{N}n}$ where the fitted parameters are 
$C_{4}$ = 0.18, $C_{5}$ = 0.12, $C_{6}$ = 0.09 and $C_{7}$ = 0.068. 
The ratio $C_{N}/B_{N}$ = 0.41 for $N$ =4, 0.43 for $N$ = 5, 0.46 for $N$ = 6 and 0.47 for $N$ = 7, 
is close to the value of 0.5 as reported in Ref.~\cite{Mth} for trapped bosons. 
\begin{figure}[hbpt]
\vspace{-10pt}
\centerline{
\hspace{-3.3mm}
\rotatebox{0}{\epsfxsize=8cm\epsfbox{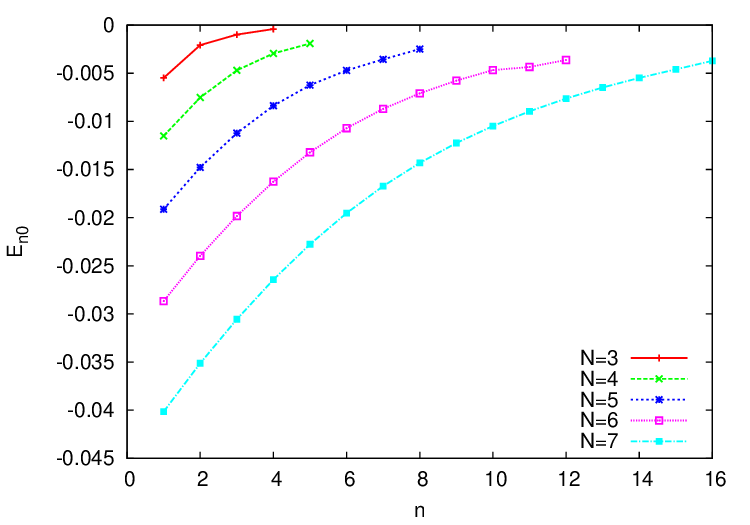}}}
\caption{Plot of energies $E_{n0}$ (in cm$^{-1}$) of van der Waals clusters of different sizes $N$ 
as a function of state number $n$ near the unitarity. Points on the curve represent the bound states.}
\label{fig.excited}
\end{figure}

\begin{figure}[hbpt]
\vspace{-10pt}
\centerline{
\hspace{-3.3mm}
\rotatebox{0}{\epsfxsize=8cm\epsfbox{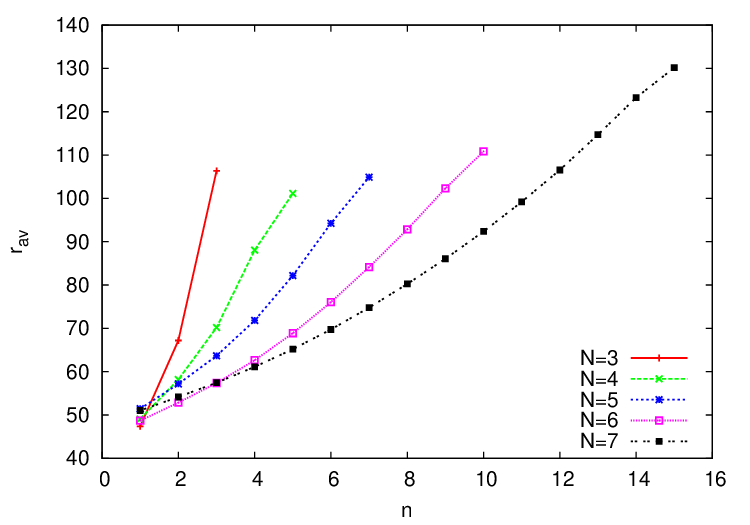}}}
\caption{Plot of the radii $r_{av}$ (in Bohr) of van der Waals cluster states of different sizes $N$ as a 
function of state number $n$ near the unitarity. Points on the curve correspond to the bound states.}
\label{fig.rav}
\end{figure}
\subsection{Structural properties and correlation}

\begin{figure}[hbpt]
\vspace{-10pt}
\centerline{
\hspace{-3.3mm}
\rotatebox{0}{\epsfxsize=8cm\epsfbox{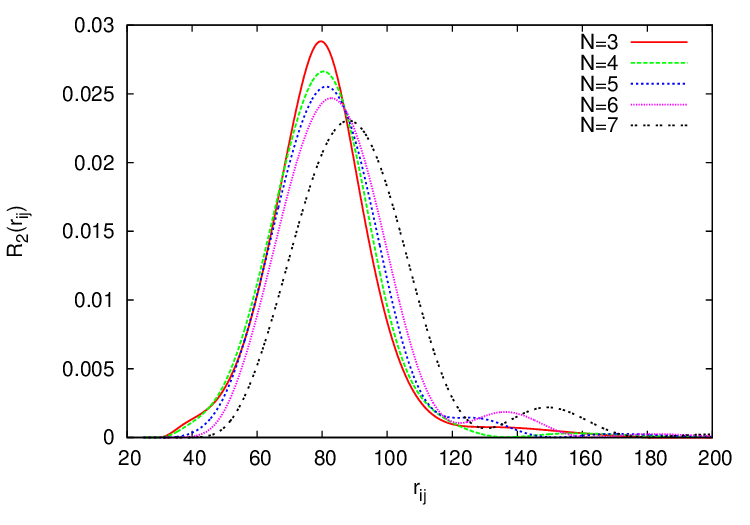}}}
\caption{Plot of pair-correlation function of van der Waals cluster of different sizes $N$ near the unitarity.}
\label{fig.correlation}
\end{figure}

\begin{figure}
  \begin{center}
    \begin{tabular}{cc}
	\resizebox{80mm}{!}{\includegraphics[angle=0]{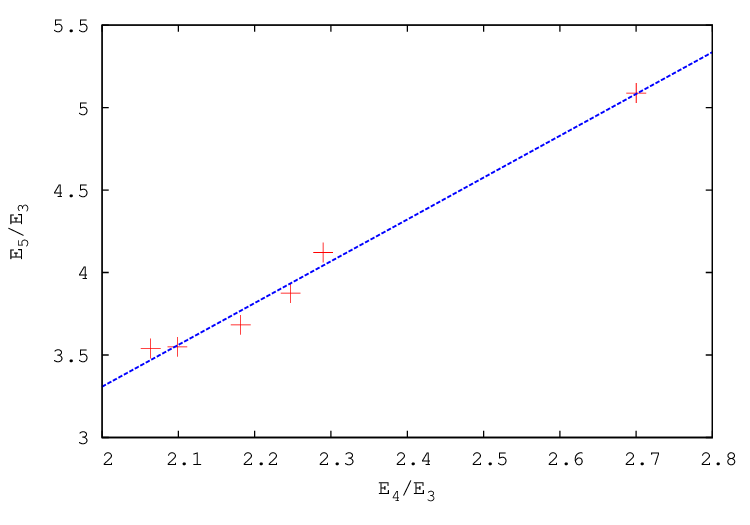}}\\  
	 (a) $N=4$ & \\
	\resizebox{80mm}{!}{\includegraphics[angle=0]{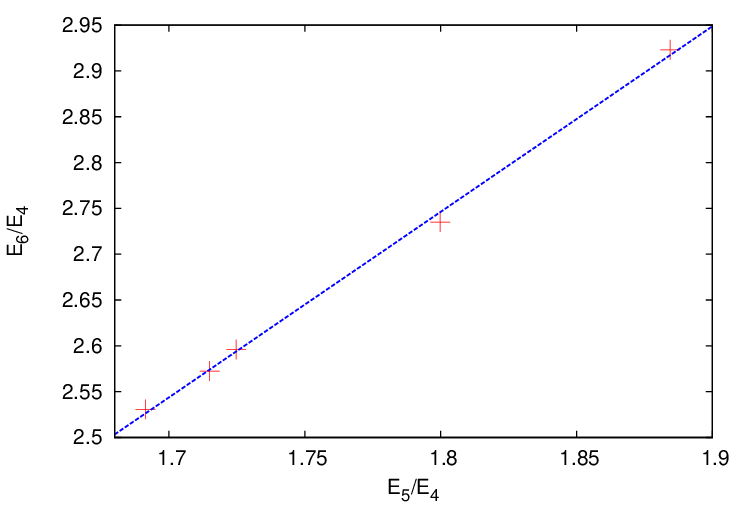}} &\\
	(b)  $N=5$ & \\
	\resizebox{80mm}{!}{\includegraphics[angle=0]{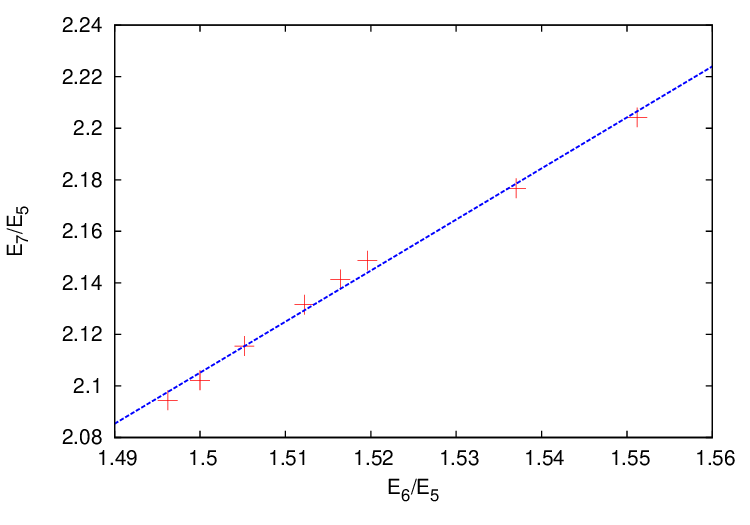}}\\  
	 (c) $N=6$ & \\
 \end{tabular}
  \end{center}
\caption{Plot of $\frac{E_{N+1}}{E_{N-1}}$ as function of $\frac{E_N}{E_{N-1}}$ for different cluster sizes $N$. 
The $+$ signs shows the numerical data and the blue dotted line represent our two-parameter linear fitting (see text) - the Tjon line.}
\label{fig.tjon}
\end{figure}

Finally we analyse the structural properties of the cluster states by calculating the pair-correlation function 
$R_2(r_{ij})$ which determines the probability of finding the $(ij)$ pair of particles at a relative separation $r_{ij}$. 
Fig.~\ref{fig.correlation} presents the pair correlation function for $N = 3-7$ at unitarity. $R_2(r_{ij})$ is considered as a more effective 
quantity in the description of structural properties as the interatomic interaction plays a crucial role. When atoms try to form clusters, 
due to the attractive part of van der Waals interaction, the short range hard core repulsion has the effect of repulsion, 
Thus $R_2(r_{ij})$ is zero for $r_{ij}$ smaller than the hard core radius $r_{c}$. We calculate $R_2(r_{ij})$ by
\begin{equation}
 R_2(r_{ij}) = \int_{\tau^{\prime \prime}} |\psi|^{2} \rd^3\tau^{\prime \prime} 
\end{equation}
where $\psi$ is the many-body wave function and the integral over the 
hypervolume excludes the integration over $r_{ij}$. The position of the maximum is shifted to larger $r_{ij}$ with increase 
in $N$ and peak height reduces. However we do not observe any structure in the correlation function. It says that the extremely 
diffuse cluster behaves just like diffuse liquid blob as observed in earlier work~\cite{Blum4}.
It is already mentioned that while the universal behaviors of the trimer are quite well understood, much less is 
known about the larger systems. In this context the investigation of correlations between energies of three and four-particle systems 
is indeed required. The earlier studies in this direction are mainly focused on the Tjon line which refers to the 
approximately linear correlation between the energies of three-nucleon and four-nucleon systems~\cite{gip,yam}. It is expected that the 
bosonic cluster energy close to the unitarity, for different cluster states should follow the generalized Tjon line. 
It says that the energies are linearly correlated to each other and a two-parameter relation is maintained. 
\begin{equation}
\frac{E_{N+1}}{E_{N-1}} =\rho_{N} + \zeta_{N} \frac{E_{N}}{E_{N-1}}
\end{equation}
In Fig.~\ref{fig.tjon}, we present the energy ratio $\frac{E_{N+1}}{E_{N-1}}$ as a function of $\frac{E_{N}}{E_{N-1}}$ 
for different cluster sizes $N$ = 4,5,6. Solid lines show linear fits of the form 
$\frac{E_{N+1}}{E_{N-1}} = \rho_{N} + \zeta_{N} \frac{E_{N}}{E_{N-1}}$. 
The fitting parameters are summarized in Table~\ref{table-tjon}. We refer the approximate linear fitting of 
the energy ratios of clusters as the generalized Tjon line. We observe that the values of the fitting parameters gradually decreases 
with increasing $N$ and this is consistent with earlier finding~\cite{Blum4}.

\begin{table}[h]
 \centering
 \caption{Values of fitting parameters of Tjon line.}
 \begin{tabular}{|l|c|c|}
 \hline
 $N$ & $\rho_{N}$ & $\zeta_{N}$ \\ \hline
 4 & $-1.76107$ & $2.5346$ \\ \hline
 5 & $-0.898113$ & $2.02464$ \\ \hline
 6 & $-0.8666535$ & $1.98111$ \\ \hline
 \end{tabular}
\label{table-tjon}
\end{table}

This definitely opens the possibilities of future investigations of how the behaviour of the generalized Tjon lines are related 
in the description of the universal properties of diffuse bosonic clusters. 

\subsection{Comparison with diffuse cluster}
The structural properties and energetics of diffuse $^{87}Rb$ clusters in three dimension has been reported earlier~\cite{Pankaj,PRA2014}. We have solved the many-particle Schr\"odinger equation by correlated PHEM and the properties of diffuse cluster have been reported for dimer scattering length and for some other specific small positive scattering length which maintains the diluteness of the cluster. The main motivation of the earlier work is to demonstrate how the cluster properties change with change in the cluster size. However in the present manuscript we are interested in small cluster but at very large negative scattering length. We basically tune the scattering length to unitary regime in search of the universal behavior. This regime is in contrast of our earlier work. The diffuse cluster is always less correlated and less complex whereas the cluster at the unitary is highly correlated and complex. Unlike the diffuse cluster with positive scattering length, where the cluster becomes less attractive with increase in scattering length, which basically makes the cluster less attractive; here we observe saturation in the ground state energy when scattering length is truly at the unitary regime.The exponential dependence of $E_{no}$ and $r_{av}$ for different cluster states of different size are completely new in the present work. To measure the correlation and its dependence on different cluster sizes, we calculate the generalized Tjon line as reported earlier. The Tjon line maintains a linear relation with two fitting parameters as observed in diffuse clusters. However the numerically fitted parameters are different. For diffuse clusters, the fitting parameters smoothly change with cluster size, however we observe significant variation in the present calculation.

\subsection{Calculation of energy levels and spectral statistics}

In our many-body picture the collective motion of the cluster is characterized by the effective potential $\omega_{0}(r)$. Thus the excited state in this potential are the states with the $l$ th surface mode and the $n$ th radial excitation, which are denoted by $E_{nl}$. Thus $n=0$ and $l=0$ correspond to the ground state and for $l \neq 0$ we get the surface modes. To calculate the higher levels with $l \neq 0$ we follow the next procedure.The weight function $\omega_l(z)$ becomes very critical for $l > 0$ and hence the numerical solution of Eq.\ref{eq.potmat}  involves a large error. We have approximated the potential matrix element for $l >0$ by the same for $l = 0$, but retaining the full hypercentrifugal repulsion of Eq.\ref{eq.cde}  which comes from the total kinetic energy operator $T$ of Eq.~(\ref{eq.Faddeev-eqn})
For small N (as we fix up $N=7$ in the present calculation), this approximation is quite good. The contribution of the hypercentrifugal term is much larger than the contribution from the potential matrix element. Thus the right side of Eq.~(\ref{eq.Faddeev-eqn}) acts as a small perturbation. Thus we take the full effect of all $l\geq 0$ from the left side of Eq.~(\ref{eq.Faddeev-eqn}) (coming from the total kinetic energy operator T). Only the effect of $l > 0$ on the small perturbation $V(r_{ij})$ is disregarded. This is also to be noted that we have extensively applied the same approximation for the calculation of several thermodynamics and condensate and statistical fluctuation of trapped bosons.\cite{r1,r2,r3}\\  

Nearest neighbour spacing distribution NNSD or $P(s)$ distribution is the most common observable which is used to study the short-range fluctuation. However to compare the statistical properties of different parts of the spectrum we need to unfold them so that the mean level density is a constant. The unfolding procedure used in our present computation involves fitting the computed energy levels $E$ to seventh order polynomial. Thus the local mean density of the unfolded spectrum becomes unity. Next we utilize the unfolded spectrum to calculate the NNSD. From the unfolded spectrum we calculate the nearest neighbour spacing as $s= E_{i+1}-E_{i}$. The NNS distribution function of a chaotic Hamiltonian is very close to that for a GOE as given by $P(s) = \frac{\pi s}{2}exp\left(-\frac{\pi s^{2}}{4} \right)$. This is also known as Wigner Dyson distribution and exhibits level repulsion between nearest neighbours ~\cite{mehta}. Whereas $P(s)= e^{-s}$ corresponds to uncorrelated spectrum and is known as Poisson statistics. Our numerical results for the lowest 30 levels are plotted in Fig. 7. We observe that $P(s)$ distribution of the van der Waals clusters closely resembles to the semi-Poisson (SP) distribution as given by $P(s) = 4 s e^{-2s}$. We observe the level repulsion at smaller values of $s$ ( $s << 1$) where $P(s)$ $\propto$ $s$ and asymptotic decay of $P(s)$ is exponential. Thus the energy levels are highly correlated which exhibits the complexity of the many-body Hamiltonian but not chaos. NNSD characterizes mainly the short-range fluctuation in the spectrum, however in order to confirm our findings on the effect of correlation on the spectral properties we investigate the long range correlation, i.e. $\Delta_{3}$ statistics. \\

We are mainly interested in the $\Delta_{3}$ statistics of Dyson and Mehta~\cite{mehta} which gives a statistical measure of the rigidity of a finite spectral level sequence. For a level sequence with a constant average level spacing, the staircase function on the average follows a straight line. Thus $\Delta_{3}$ statistics gives a measure of the size of fluctuations of the staircase function around a best fit straight line. It is determined as 
\begin{equation}
 \Delta_{3} = \big \langle min_{(a_l,b_l)}\int_{E-\frac{L}{2}}^{E+\frac{L}{2}}[n(E)-a_l-b_l E]^2 dE \big \rangle
\end{equation},
while $a_l$ and $b_l$ are two constants for the least-square fit and $n(E)$ is a step-function with mean slope of one.



It is customary to use the average values of $\Delta_{3}(L)$, i.e.   $<\Delta_{3}(L)>$. For uncorrelated Poisson spectrum $<\Delta_{3}(L)>$ $\propto$ $L$. whereas for Wigner spectrum $<\Delta_{3}(L)>$  $\propto$ $log L$. Our numerical results plotted in Fig.8, it again confirms semi-Poisson distribution.

\begin{figure}[hbpt]
\vspace{-10pt}
\centerline{
\hspace{-3.3mm}
\rotatebox{0}{\epsfxsize=8cm\epsfbox{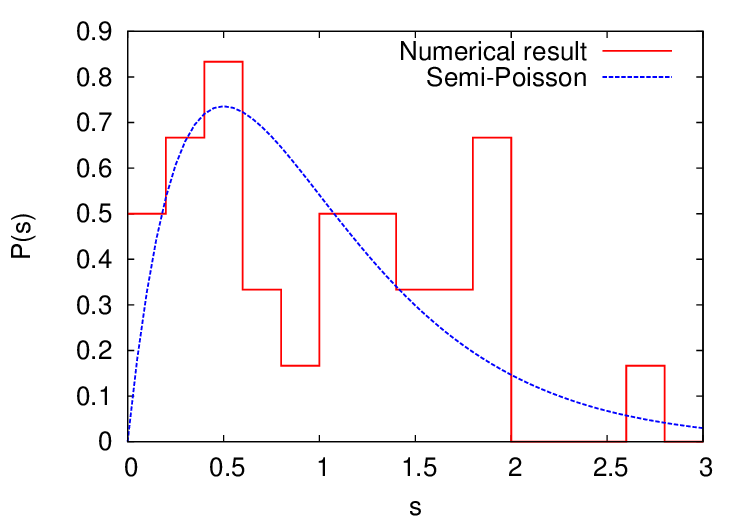}}}
\caption{Numerical results of nearest neighbour level spacing distribution $P(s)$ for $N=7$. Comparison with semi-Poisson distribution is also presented.}
\label{fig.excited}
\end{figure}

\begin{figure}[hbpt]
\vspace{-10pt}
\centerline{
\hspace{-3.3mm}
\rotatebox{0}{\epsfxsize=8cm\epsfbox{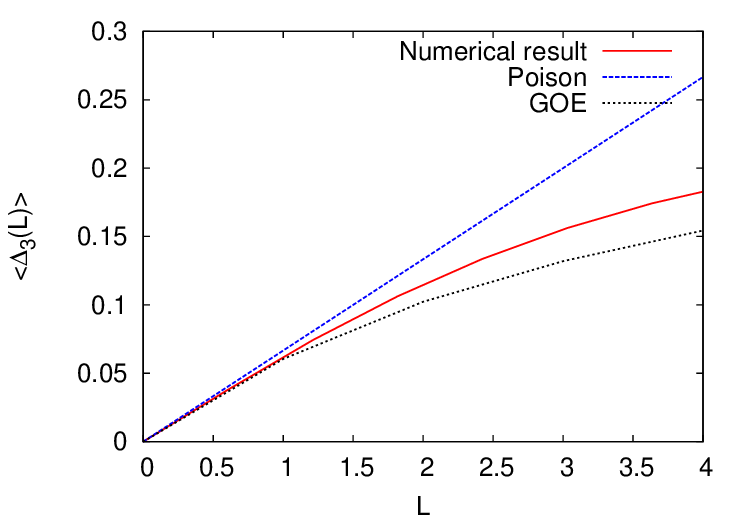}}}
\caption{Numerical results of spectral average $<\Delta_{3}(L)>$ for $N=7$}
\label{fig.excited}
\end{figure}

However distribution of the ratio of consecutive level spacing $P(r)$ and the related averages are presently utilized as  the most useful statistical measures of complex many-body system \cite{r4,r5,r6,r7}. As the calculation of $P(r)$ does not require any unfolding of the energy spectrum, it is independent of the  form of the  density of energy levels. For the ordered set of eigenvalues $E_n$, the nearest level spectrum is $s_n = E_{n+1}- E_n$ and the ratio of two consecutive level spacing is $r_n = \frac{s_{n+1}}{s_n}$. The probability of consecutive level spacing is $P(r)$. For the completely integrable system $P(r)$ follows Poisson distribution which is given by,

\begin{figure}[hbpt]
  \begin{center}
    \begin{tabular}{cc}

      \resizebox{47mm}{!}{\includegraphics[angle=0,width=5mm]{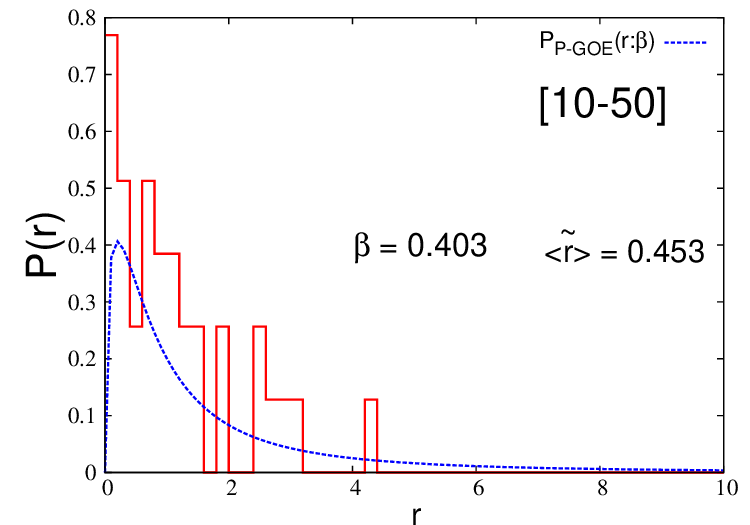}}&        
      \resizebox{47mm}{!}{\includegraphics[angle=0]{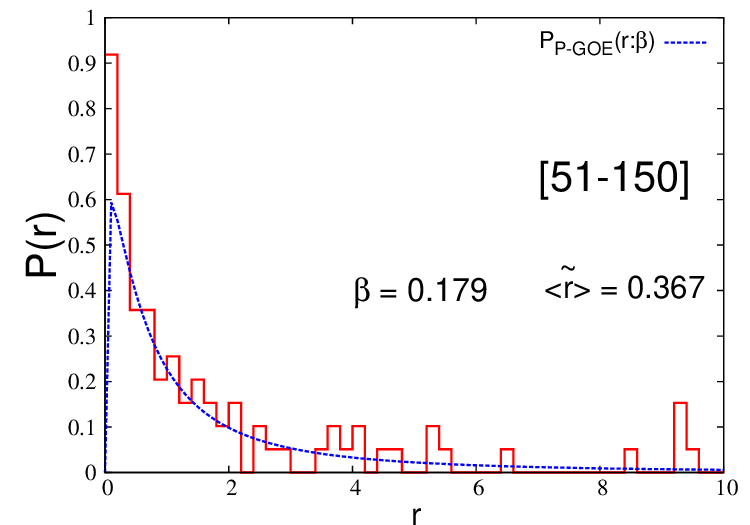}}\\
      (a)&
      (b)\\
      \resizebox{47mm}{!}{\includegraphics[angle=0,width=50mm]{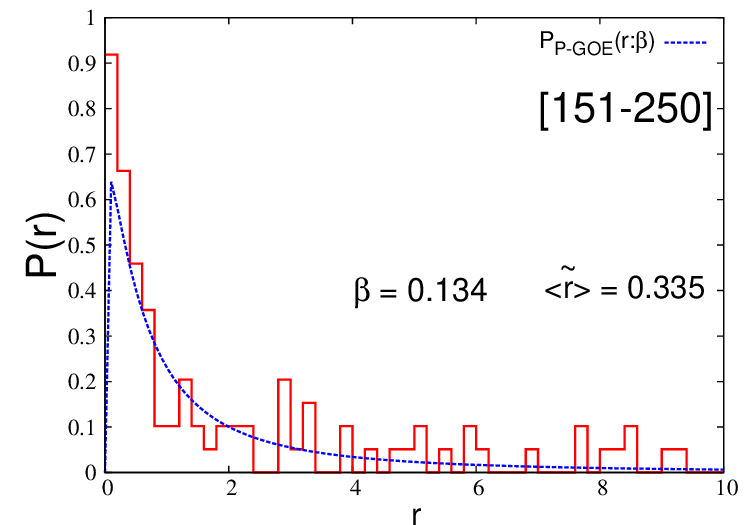}}&         
      \resizebox{47mm}{!}{\includegraphics[angle=0]{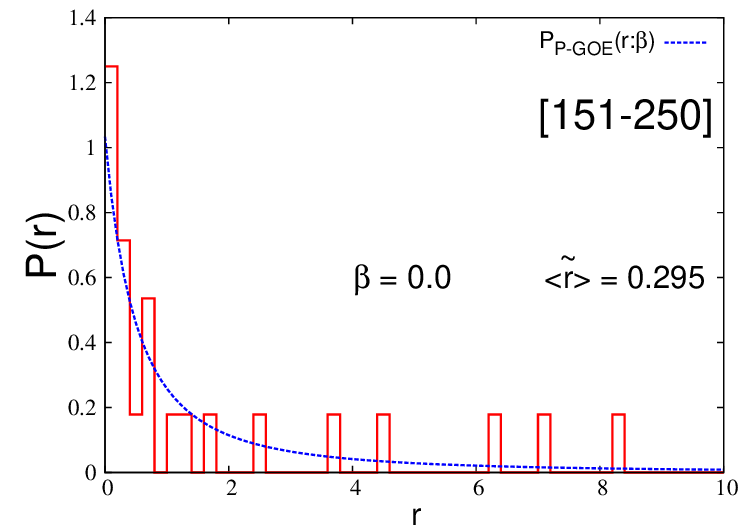}}\\
      (c)&
      (d)\\	

       \end{tabular}
  \end{center}
  \caption {Plot of $P(r)$ distributions for different energy level window}
  \label{fig.pr}
\end{figure}

\[ P_P(r) = \frac{1}{(1+r)^2}\]
whereas for chaotic region, it follows Wigner distribution,
\[P_W(r) = \frac{27}{8}\frac{r+r^2}{(1+r+r^2)}\]
 It is also possible to consider $\tilde{r}$ given by,
\[\tilde{r} = \frac{min(S_n,S_{n-1})}{max(S_n,S_{n-1})} = min (r_n,\frac{1}{r_n})\]
The average value of $\tilde{r}$, i.e. $<\tilde{r}> = 0.56$ for GOE and $0.386$ for Poisson.

In Fig.\ref{fig.pr},we plot $P(r)$ distribution for several energy level in different panel. We fit the histogram by utilizing the interpolation formula from Poisson to GOE is given by,
\begin{equation}
P_{P-GOE}(r:\beta) = \frac{1}{Z_\beta} \frac{(r+r^2)^\beta}{(1+(2-\beta)r+r^2)^{1+\frac{3}{2}\beta}}
\end{equation}

\begin{figure}[hbpt]
\vspace{-10pt}
\centerline{
\hspace{-3.3mm}
\rotatebox{0}{\epsfxsize=8cm\epsfbox{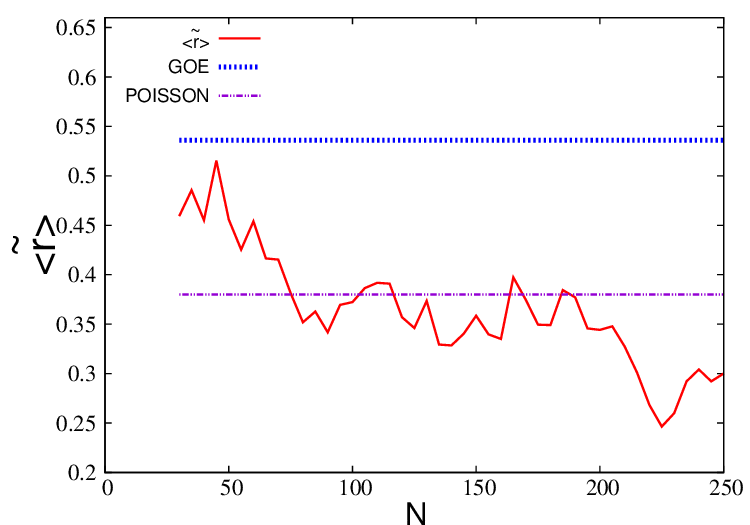}}}
\caption{Average value of $\tilde{r}$ as a function of energy levels.}
\label{fig.prn}
\end{figure}

$\beta = 0$ gives Poisson and $\beta = 1$ gives GOE. The parameter $Z_\beta$ is obtained using the condition $\int_0^\infty P(r) dr = 1 $.\\
In panel(a), we plot the results for 10-50 levels. We do not use lower levels which shows large fluctuation. The fitted $\beta$ parameter is 0.403 which lies between GOE and Poisson and the calculated $<\tilde{r}> = 0.453$. In panel $(b)-(d)$, we plot the same for different windows of energy levels. The fitting parameter $\beta$ gradually goes to zero and $<\tilde{r}>$ reaches its Poisson limit as we move to higher levels which shows smooth transition from semi-Poisson to Poisson distribution.

In Fig.\ref{fig.prn}, we plot $<\tilde{r}>$ as a function of number of energy levels. The $<\tilde{r}>$ is below the GOE value initially and gradually reaches the Poisson limit with increase in number of levels.


Further we use a $\chi^2$ test to measure the distance from the numerical result to the theoretical prediction. $\chi^2$ is defined as,
\begin{equation}
\chi_{\alpha}^2 = log_{10} \left\lbrace  \int_0^\infty dS \{ P_\alpha(S) - P(S)\} \right\rbrace 
\end{equation}
where $\alpha$ stands for Poisson or Wigner Dyson or Semi-Poisson.

\begin{figure}[hbpt]
	\begin{center}
    \begin{tabular}{cc}
	
      \resizebox{43mm}{!}{\includegraphics[angle=0]{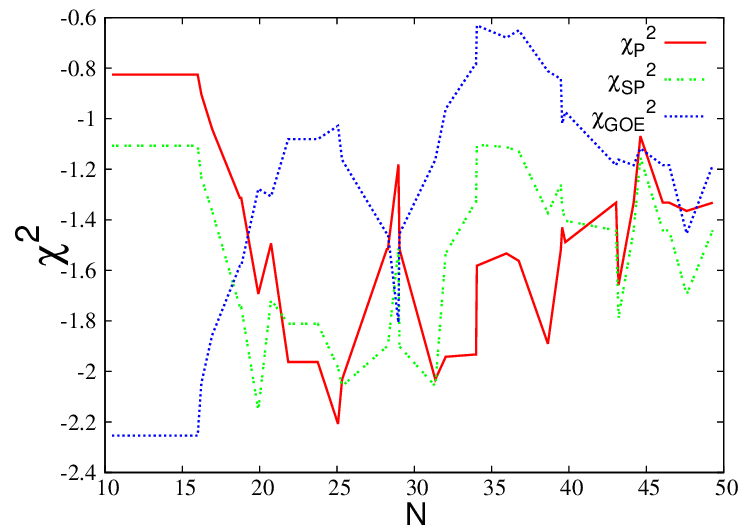}}&         
      \resizebox{42mm}{!}{\includegraphics[angle=0]{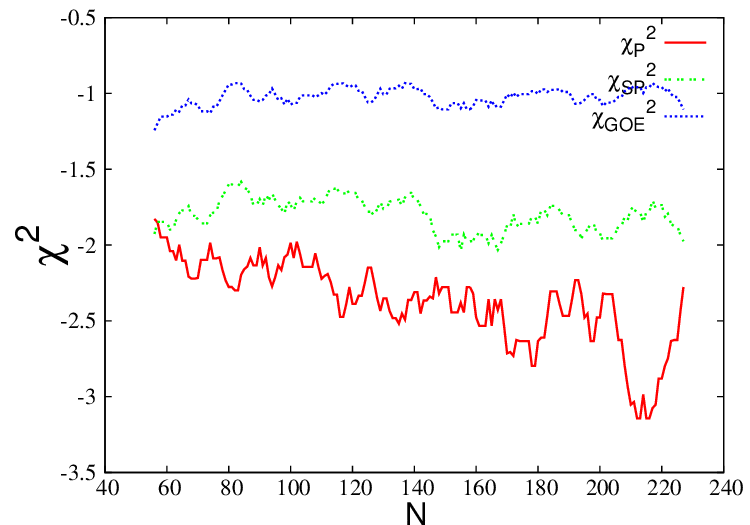}}\\
      (a)&
      (b)\\ 	
	
       \end{tabular}
       \end{center}
       \caption {$\chi^2$ statistical tests for lower levels (Panel a) and for higher levels (Panel b)}
       \label{fig.chi1}

\end{figure}

In Fig.\ref{fig.chi1}, we plot it as a function of energy levels. From fig.\ref{fig.chi1}(a), we observe lot of variation in $\chi^2$ plot. $\chi_{GOE}^2$ is always highest expect only for first few levels. Where as $\chi_{SP}^2$ and $\chi_{P}^2$ intersects at various points. Where as in Fig.\ref{fig.chi1}(b)(for the higher levels), $\chi_{P}^2$ is always the lowest which satisfies our earlier observation made in Fig.\ref{fig.pr}. So, analysing all possible statistical measures in the present calculation, we conclude a smooth transition from $SP$ to Poisson statistics. However with increase in cluster size, the correlation and complexity gradually builds in. One may expect results closer to GOE at least for lower levels and possibility of chaos is not ruled out. 
\section{Conclusion}

The physics of weakly bound few-body systems and their universal behavior near the unitary is a challenging research area in recent days. 
The recent experimental achievement of ultracold Bose gases has renewed the interest in universal few-body physics. 
The theoretical study of three-dimensional bosonic cluster with more than three particles is also challenging and the numerical treatment 
becomes complicated with $N > 3$. The cluster is weakly bound as the kinetic and potential energy nearly cancel. 
It needs to include interatomic correlation. In the present study we utilize two-body correlated basis function for the study of $N$-boson systems. 
Use of realistic van der Waals potential presents the actual feature of such delicate systems. We calculate the energy spectrum of $N$-body cluster 
with $N$ upto 7 atoms. At large scattering length, which is much larger than the range of interaction, the ground state energy of $N$-body cluster 
shows universal behaviour. \\
The spatially extended energy states exhibit the exponential dependence on the state number. 
We also calculate the r.m.s radii of the spatially extended systems and also shows their exponential dependence on the state number. 
Calculation of two-body pair correlation exhibit the expected feature and does not show any structure. 
It says that the weakly interacting cluster behaves just like a single quantum stuff. We also calculate the energy correlation between two clusters 
differing by one atom and shows that they maintain a two parameter linear relation. We refer the Tjon line as the characteristic of universal behaviour 
of bosonic cluster. We study the nearest-neighbor spacing distribution and the spectral rigidity by unfolding the spectrum. We assume that the spectrum exhibits semi-Poisson statistics both in the calculation of short-range fluctuation and long-range correlation and in other statistical measures. Although in the present calculation with small cluster having only $7$ atoms, we do not see any signature of chaos, however for larger cluster size the possibility of chaos can not be ruled out. 
\begin{center}
{\large{\bf{Acknowledgements}}}
\end{center}
BC would like to thank the University of South Africa (UNISA) for the financial support of visit where part of work was done. 
BC also acknowledges financial support of DST (India) under the research grant SR/S2/CMP-0126/2012 and also WBDST for the research grant 1211(Sanc.)/ST/P/S\&T/4G-1/2012.

\end{document}